\documentclass[letterpaper,twocolumn,prl,aps,superscriptaddress,amsmath,amssymb,floatfix]{revtex4-1}
\usepackage{natbib}
\usepackage{mathptmx}
\usepackage[latin9]{inputenc}
\usepackage{color}
\usepackage{verbatim}
\usepackage{braket}
\usepackage{amssymb,amsmath,amssymb,amsthm,mathrsfs}
\usepackage{graphicx}
\usepackage{esint}
\usepackage{siunitx}
\usepackage{soul}
\usepackage[unicode=true,
 bookmarks=true,bookmarksnumbered=false,bookmarksopen=false,
 breaklinks=false,pdfborder={0 0 1},backref=false,colorlinks=true]
 {hyperref}
\hypersetup{
 linkcolor=magenta,urlcolor=blue,citecolor=blue,pdfstartview={FitH},hyperfootnotes=false}

\makeatletter

\usepackage{textcomp}
\usepackage{epstopdf}

\pdfpageheight\paperheight
\pdfpagewidth\paperwidth

\@ifundefined{textcolor}{}{%
 \definecolor{BLACK}{gray}{0}
 \definecolor{WHITE}{gray}{1}
 \definecolor{RED}{rgb}{1,0,0}
 \definecolor{GREEN}{rgb}{0,1,0}
 \definecolor{BLUE}{rgb}{0,0,1}
 \definecolor{CYAN}{cmyk}{1,0,0,0}
 \definecolor{MAGENTA}{cmyk}{0,1,0,0}
 \definecolor{YELLOW}{cmyk}{0,0,1,0}
}

\usepackage{xcolor}
\usepackage{soul}
\definecolor{blue}{rgb}{0,0,1}
\definecolor{red}{rgb}{1,0,0}
\definecolor{green}{rgb}{0,1,0}

\usepackage{soul}

\makeatother

\begin{document}

\affiliation{Laboratory of Quantum Information, University of Science and Technology of China, Hefei, Anhui 230026, China}
\affiliation{Center for Quantum Information, Institute for Interdisciplinary Information Sciences, Tsinghua University, Beijing 100084, China}
\affiliation{Hefei National Laboratory, Hefei 230088, China}

\title{Fault-tolerant preparation of arbitrary logical states in the cat code}

\author{Zi-Jie Chen}
\thanks{These authors contributed equally to this work.}
\affiliation{Laboratory of Quantum Information, University of Science and Technology of China, Hefei, Anhui 230026, China}

\author{Weizhou~Cai}
\thanks{These authors contributed equally to this work.}
\affiliation{Laboratory of Quantum Information, University of Science and Technology of China, Hefei, Anhui 230026, China}

\author{Liang-Xu Xie}
\affiliation{Laboratory of Quantum Information, University of Science and Technology of China, Hefei, Anhui 230026, China}

\author{Qing-Xuan Jie}
\affiliation{Laboratory of Quantum Information, University of Science and Technology of China, Hefei, Anhui 230026, China}

\author{Xu-Bo Zou}
\affiliation{Laboratory of Quantum Information, University of Science and Technology of China, Hefei, Anhui 230026, China}
\affiliation{Hefei National Laboratory, Hefei 230088, China}

\author{Guang-Can Guo}
\affiliation{Laboratory of Quantum Information, University of Science and Technology of China, Hefei, Anhui 230026, China}
\affiliation{Hefei National Laboratory, Hefei 230088, China}

\author{Luyan~Sun}
\email{luyansun@tsinghua.edu.cn}
\affiliation{Center for Quantum Information, Institute for Interdisciplinary Information Sciences, Tsinghua University, Beijing 100084, China}
\affiliation{Hefei National Laboratory, Hefei 230088, China}

\author{Chang-Ling Zou}
\email{clzou321@ustc.edu.cn}
\affiliation{Laboratory of Quantum Information, University of Science and Technology of China, Hefei, Anhui 230026, China}
\affiliation{Hefei National Laboratory, Hefei 230088, China}


\begin{abstract}
Preparing high-fidelity logical states is a central challenge in fault-tolerant quantum computing, yet existing approaches struggle to balance control complexity against resource overhead. Here, we present a complete framework for the fault-tolerant preparation of arbitrary logical states encoded in the four-legged cat code. This framework is engineered to suppress the dominant incoherent errors, including excitation decay and dephasing in both the bosonic mode and the ancilla via error detection. Numerical simulations with experimentally realistic parameters on a 3D superconducting cavity platform yield logical infidelities on the order of $10^{-4}$. A scaling analysis confirms that the logical error rate grows nearly quadratically with the physical error rate, confirming that all first-order errors are fully suppressed. Our protocol is compatible with current hardware and is scalable to multiple bosonic modes, providing a resource-efficient foundation for magic state preparation and higher-level concatenated quantum error correction.
\end{abstract}

\maketitle




\noindent \textit{Introduction.-} Efficient, high-fidelity state preparation is the cornerstone of quantum computing~\cite{campbell2017roads}. First, high-quality magic states are indispensable resources for achieving universal gate sets via state injection~\cite{nielsen2010quantum}, yet their preparation often dominates architectural overheads, consuming up to $90\%$ of resources in certain proposals~\cite{fowler2012surface}. Furthermore, the initialization of logical states is necessary in standard quantum error correction (QEC) schemes~\cite{gottesman2009introduction}, such as those proposed by Shor~\cite{shor1996fault}, Steane~\cite{steane1997active}, and Knill~\cite{knill2004fault}. Consequently, developing efficient protocols to prepare these states with high fidelity is crucial for the practical realization of universal quantum computation.

Existing approaches face a fundamental dilemma in balancing control complexity with resource overhead. Quantum optimal control methods, such as GRAPE~\cite{khaneja2005optimal,chen2025robust}, have been successfully demonstrated in small-scale systems, but they suffer from intractable optimization complexity as system size increases. Conversely, QEC methods for state preparation offer scalability but encounter distinct limitations: magic state distillation~\cite{bravyi2005universal} is resource-intensive and typically restricted to specific states like T-states, lacking the flexibility for direct arbitrary state preparation. While flag-qubit-based protocols~\cite{chao2018quantum,chamberland2020very} can suppress errors with reduced hardware resources, they impose demanding connectivity constraints that remain challenging for many hardware architectures.

Recently developed bosonic error correction~\cite{cai2021bosonic,ma2021quantum} provides a novel approach to error suppression by directly leveraging hardware properties and the error-correction capabilities of bosonic codes. This strategy has successfully enabled hardware-efficient operations, including error-transparent~\cite{ma2020error}, path-independent~\cite{ma2020path,reinhold2020error}, and error-detectable gates~\cite{tsunoda2023error}. Despite these advances, this strategy has not yet been fully realized for the preparation of arbitrary initial states, with all dominant errors being  suppressed.

In this Letter, we address these challenges by establishing a complete fault-tolerant (FT) framework for arbitrary state preparation using the bosonic cat code based on error detection. Our protocol engineers interactions to map dominant first-order errors to a detectable ancilla state, including spontaneous decay and dephasing in both the bosonic mode and the ancilla. By post-selecting against these heralded events, we effectively suppress logical error rates without complex hardware overhead. Numerical simulations using realistic experimental parameters demonstrate that our protocol can prepare arbitrary logical states with infidelities on the order of $10^{-4}$. Crucially, a scaling analysis reveals that the logical error rate scales almost quadratically with the physical error rate, providing direct evidence that first-order errors are successfully suppressed in our protocol.

\begin{figure}[t]
\begin{centering}
\includegraphics[scale=0.7]{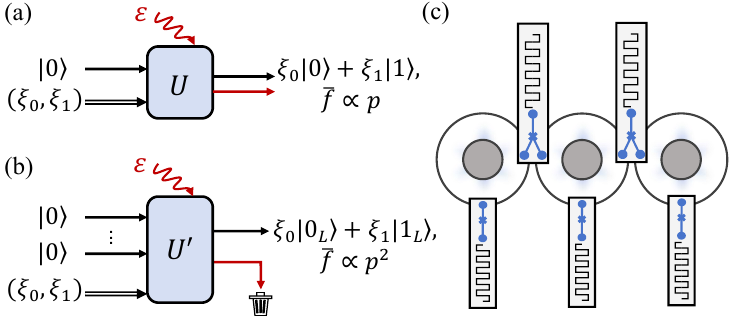}
\par\end{centering}
\caption{\label{Fig:Schematic}{Arbitrary state preparation with error-detectable quantum operation. }({a}) Initialization of a target state $\xi_0\ket{0}+\xi_1\ket{1}$ through a noisy evolution, where the infidelity $\overline{f}$ of the final state is proportional to the physical error rate $p$. ({b}) Initialization of a logical target state $\xi_0\ket{0_L}+\xi_1\ket{1_L}$ through a noisy error-detectable operation, where the first order error can be suppressed through post-selection. ({c}) Schematic of the 3D superconducting cavities (circles) platform that can realize error-detectable operations. The three chips at the bottom correspond to ancillary qubits coupled to individual cavities, while the top two chips represent ancillas coupled to both adjacent cavities.}
\end{figure}

\noindent\textit{Protocol and System.-}  Figure~\ref{Fig:Schematic}(a) illustrates a standard state preparation protocol, which is vulnerable to environmental noise. Lacking a mechanism to distinguish these errors, it inevitably results in an imperfect final state with infidelity $\overline{f}\propto p$, where $p$ is the physical error rate. In contrast, Fig.~\ref{Fig:Schematic}(b) presents our protocol based on error-detectable operations. By introducing ancillary qubits which provide extra degrees of freedom for redundancy, we engineer the system-ancilla interactions so that dominant error events leave a detectable signature in the ancilla state. Consequently, post-selecting the ancilla in particular outcomes effectively filters out these error-corrupted system states, suppressing the final infidelity to $\overline{f}\propto p^2$ or even higher orders. By constructing a universal set of such error-detectable operations for QEC code, this scheme enables the fault-tolerant (FT) preparation of arbitrary logical states.

We consider the 3D superconducting circuit quantum electrodynamics (QED) platform sketched in Fig.~\ref{Fig:Schematic}(c), in which high-quality-factor microwave cavities (serving as bosonic modes for QEC encodings) are dispersively coupled to three-level ancilla qubits with levels $\ket{g}$, $\ket{e}$, and $\ket{f}$. The dispersive Hamiltonian for a single mode (denoted by bosonic operator $a$) reads
\begin{align}
    H_{0}=\chi_{f} \ket{f} \bra{f}\otimes a^\dagger a +\chi_{e}  \ket{e} \bra{e} \otimes a^\dagger a,
\end{align}
where $\chi_{f,e}$ are the cross-Kerr interaction coefficients, giving rise to photon-number-dependent frequency shifts on the ancilla transitions. The control Hamiltonian $H_\mathrm{c}$ comprises a coherent drive on the cavity for displacement operations $D(\alpha)$ with amplitude $\alpha$ and a drive coupling the ancilla two-photon transition $\ket{g}\leftrightarrow\ket{f}$, while $\ket{e}$ is never intentionally populated and is reserved exclusively for error detection. The dominant hardware error channels are
\begin{equation}
 \epsilon_\mathrm{h}\in \{   a,a^\dagger a ,\ket{e}\bra{f}, \sum_{m\in\{e,f\}}\epsilon_m  \ket{m}\bra{m} \},
    \label{Equ:ErrorSet}
\end{equation}
representing cavity photon loss, dephasing, and population relaxation and the dephasing of the ancilla, respectively.

We encode the logical qubit in the four-legged cat code~\cite{leghtas2013hardware}:
\begin{equation}\label{Equ:CatCode}
    \ket{0_\mathrm{c}/1_\mathrm{c}}\propto \ket{\alpha}+\ket{-\alpha}\pm\ket{i\alpha}\pm\ket{-i\alpha},
\end{equation}
which are superpositions of four coherent states, with $\ket{\alpha}=D(\alpha)\ket{\mathrm{vac}}$ representing a coherent state and $\ket{\mathrm{vac}}$ denoting the vacuum. We note that when the amplitude becomes sufficiently large ($|\alpha|^2\gg1$), these four components are approximately orthogonal, and the cat code encoding becomes significantly less sensitive to the bosonic dephasing error $a^\dagger a$ and the mean photon number of the code words are approximately equal.

\begin{figure}[t]
\begin{centering}
\includegraphics[scale=0.7]{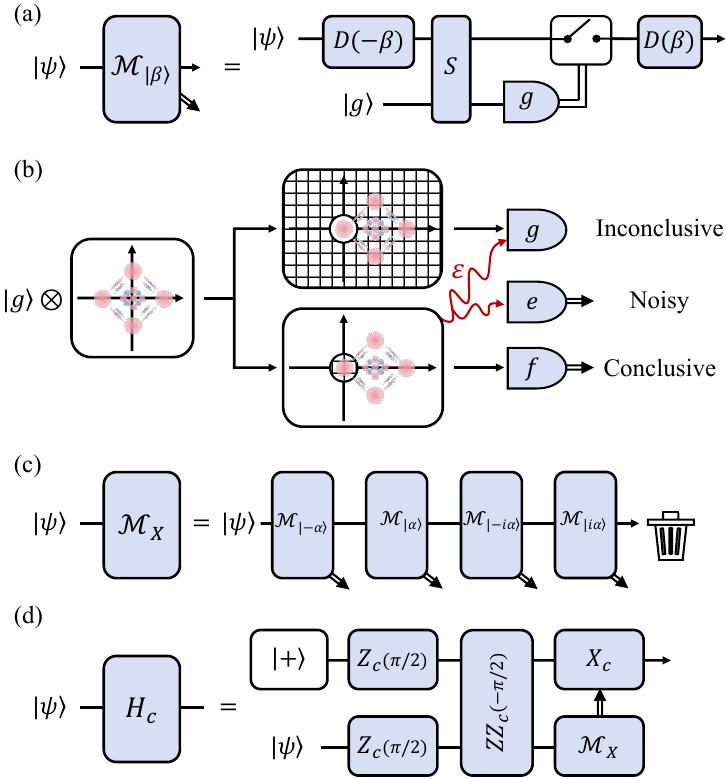}
\par\end{centering}
\caption{{Error-detectable Pauli-$X_\mathrm{c}$ measurement.}
({a}) Circuit for the unambiguous quantum state discrimination of a coherent state component $|\beta\rangle$, employing a selective bit-flip
gate $S$.
({b}) Phase-space schematic of the state discrimination for the $|-\alpha\rangle$ component of the cat code using the circuit in ({a}). When the ancilla is measured in $|f\rangle$, the measurement terminates, yielding an accurate conclusive result. However, an outcome of $|e\rangle$ indicates an error occurred, leading to the discard of the state and termination. A $|g\rangle$ outcome implies either that the system did not collapse into the target component or that it suffered from error, thus yielding an inconclusive result requiring subsequent measurements.
({c}) Implementation of the Pauli-$X_\mathrm{c}$ measurement $\mathcal{M}_X$ via sequential measurement of the four coherent state components of the cat code. The sequence terminates with an $|f\rangle$ outcome. If all four stages yield $|g\rangle$, the measurement fails and state is discarded.
({d}) Implementation of $H_\mathrm{c}$ gate through the error-detectable $Z_\mathrm{c}(\theta)$, $ZZ_\mathrm{c}(\theta)$, and $\mathcal{M}_X$.
}
\label{Fig:UniversalCircuit}
\end{figure}

Expanding the code words in the Fock basis, the interference among four coherent states with different phase factors exhibit populations only on photon numbers ($n$) of certain parities: $n\equiv0/2$ (mod 4) for $\ket{0_\mathrm{c}/1_\mathrm{c}}$. Therefore, the logical qubits are protected from the single photon loss error via parity measurement $\mathcal{M}_P$, which is enabled by the dispersive coupling through a phase accumulation  $C_\phi=\mathrm{exp}(-i\phi\ket{f}\bra{f}\otimes a^\dagger a)$ with $\phi=\pi$. A $|g\rangle$ outcome projects the bosonic mode into the even-parity subspace $\Pi_\mathrm{even}=\sum_n \ket{2n}\bra{2n}$, while an $|f\rangle$ outcome implements  $  \Pi_\mathrm{odd}=\sum_n \ket{2n+1}\bra{2n+1}$. This operation is naturally error-detectable against all dominant hardware errors in Eq.~(\ref{Equ:ErrorSet}) (See \cite{SM} for details). The parity measurement immediately enables the FT preparation of a logical qubit in $X$-basis eigenstates with built-in error detection: applying $\mathcal{M}_P$ to a coherent state $|\alpha\rangle$ {or $\ket{i\alpha}$} and post-selecting on the outcome yields $\ket{+_\mathrm{c}} \propto \ket{\alpha} + \ket{-\alpha}$ or $\ket{-_\mathrm{c}} \propto \ket{i\alpha} + \ket{-i\alpha}$.

However, the existing tools for the cat code have limitations in the FT preparation of arbitrary states. The error-detectable operations are limited to the $n$-dependent phase accumulations and do not change the photon number, thus the rotations of logical qubit along $X$- or $Y$-axis that move populations from $n\equiv0$ (mod 4) to $n\equiv2$ (mod 4) are inhibited.

\smallskip{}
\noindent\textit{Error-detectable Pauli-$X_\mathrm{c}$ measurement.-}
To overcome the limitations based on photon-number preserving interactions, we propose the FT coherent state discrimination $\mathcal{M}_{|\beta\rangle}$, as depicted by the quantum circuit in Fig.~\ref{Fig:UniversalCircuit}(a). The discrimination can determine whether the mode is in a particular coherent-state component $\ket{\beta}$ with $\beta\in\{\pm\alpha,\pm i\alpha\}$ by three steps of operations:

\begin{enumerate}
    \item \textit{Displacement to the origin $\ket{\mathrm{vac}}$}. A displacement $D(-\beta)$ translates the phase space so that the target component $\ket{\beta}$ is mapped to $\ket{\mathrm{vac}}$, which is equivalent to $\ket{0}$ in Fock basis, while other orthogonal coherent state components are shifted to points far from the origin.
    \item \textit{Photon-number selective ancilla flip}. Employing the dispersive interaction, an operation
    \begin{equation}    S=i\sigma_x\otimes(|0\rangle\langle0|+|1\rangle\langle1|)+I_q\otimes\sum_{n\ge2}|n\rangle\langle n|
    \end{equation}
    is applied, where $\sigma_x$ is the Pauli operator within the ancilla $\{|g\rangle, |f\rangle\}$ subspace. $S$ is engineered to flip the ancilla from $|g\rangle$ to $|f\rangle$ at target Fock state $\ket{0}$. We note that $S$ is also engineered to act on $|1\rangle$, ensuring resilience against bosonic dephasing errors.
    \item \textit{Reverse displacement}.  A displacement $D(\beta)$ restores the cavity to its original phase-space configuration.
\end{enumerate}

The phase-space dynamics of the $\mathcal{M}_{|\beta\rangle}$ measurement (with $\beta=-\alpha$) is visualized in Fig.~\ref{Fig:UniversalCircuit}(b), showing the evolution of the Wigner function for a cat code. The potential outputs of the measurement are $\ket{g}$, $\ket{e}$, and $\ket{f}$, presenting an unambiguous quantum state discrimination operation~\cite{acin2001statistical,cai2024unambiguous}: the output at $\ket{f}$ gives a conclusive result of input state being $\ket{\beta}$. Such detection is FT against the dominant hardware errors described in Eq.~(\ref{Equ:ErrorSet}): (i) Ancilla decay $\ket{e}\bra{f}$ can be detected through the output at $\ket{e}$. (ii) Ancilla dephasing would suppress the transition from $\ket{g}$ to $\ket{f}$, but will not induce a false transition. (iii) Cavity photon loss during the operation will shift the effective displacement, reducing the probability of a successful ancilla flip but not creating a false positive. (iv) Cavity dephasing is tolerated as we include the Fock state $\ket{1}$ in $S$ as well. In every case, the error either suppresses a correct detection (contributing to the inconclusive channel) or flags itself through the $\ket{e}$ outcome, and no first-order error produces a false conclusive result.

For the four-legged cat code [Eq.~(\ref{Equ:CatCode})], the $\mathcal{M}_{X}$ measurement can be realized through the sequential discrimination of the four coherent state components. As depicted in Fig.~\ref{Fig:UniversalCircuit}(c), the measurement proceeds sequentially by checking for the $|-\alpha\rangle$, $|\alpha\rangle$, $|-i\alpha\rangle$, and $|i\alpha\rangle$ components. In the limit of $|\alpha| \gg 1$, the  four coherent state components of the cat code become almost orthogonal, and it ensures that the probability of the ancilla being flipped by non-target components is exponentially suppressed. More explicitly, the consecutive outputs of ``fggg" and ``gfgg" indicate the output at $\ket{+_\mathrm{c}}$ while ``ggfg" and ``gggf" indicate the output at $\ket{-_\mathrm{c}}$, and all other outputs are discarded. The same procedure can be extended to FT discrimination of joint coherent states in multiple modes, and generalizes to the joint Pauli-$XX$ measurement $\mathcal{M}_{XX}$ in an architecture where a single ancilla is coupled to two cavities (see \cite{SM} for details).

\smallskip{}
\noindent\textit{Error-detectable gate set.-}
With the $\mathcal{M}_X$ measurement established, we can realize the Hadamard gate $H_\mathrm{c}$ via a one-bit teleportation protocol, as depicted in the circuit in Fig.~\ref{Fig:UniversalCircuit}(d). This scheme relies on the injection of an ancillary state $\ket{+_\mathrm{c}}$, which is prepared in a FT manner through $\mathcal{M}_p$. A Pauli-$X_\mathrm{c}$ correction on the output state should be implemented according to the measurement outcome of the $\mathcal{M}_X$ measurement. While this correction could physically be realized through a phase $e^{i\pi/2a^\dagger a}$ on the mode, we can instead employ standard Pauli frame tracking for convenience.

Prior works~\cite{rosenblum2018fault,sun2014tracking} have established a set of error-detectable operations for the cat code, including parity measurement $\mathcal{M}_P$, single qubit arbitrary-angle Pauli-$Z_\mathrm{c}$ rotation $Z_\mathrm{c}(\theta)=\mathrm{exp}(-i\frac{\theta}{2} Z_\mathrm{c})$, and two-qubit $ZZ_\mathrm{c}$ rotation $ZZ_\mathrm{c}(\theta)=\mathrm{exp}(-i\frac{\theta}{2} Z_{\mathrm{c},1} Z_{\mathrm{c},2})$~\cite{tsunoda2023error}. All three are built from the same dispersive cross-Kerr interaction and tolerate all dominant hardware error channels (see \cite{SM} for details).

Therefore, we complete the universal set of error-detectable operations for the cat code as
\begin{align}\label{Equ:UniversalGate}
    \{ \mathcal{M}_{P}, Z_\mathrm{c}(\theta), ZZ_\mathrm{c}(\theta), \mathcal{M}_{X},H_\mathrm{c} \},
\end{align}
and it serves as a sufficient toolkit for the preparation of arbitrary states. Compared with previous schemes~\cite{xu2024fault,guillaud2019repetition,puri2020bias} for constructing fault-tolerant operations with bosonic codes, state preparation presents two features that make this error-detectable framework particularly powerful. First, state preparation circuits are inherently shallow, eliminating the need for additional hardware overheads to autonomously stabilize the bosonic code, such as engineered dissipation or high-order non-linearity~\cite{guillaud2019repetition,puri2020bias}. Second, post-selection is free: a failed preparation is simply discarded and repeated at negligible cost. For example, it is not necessary to implement generalized path-independent gates~\cite{xu2024fault}, as error-detectable gates are sufficient. Consequently, this framework allows for a more resource-efficient suppression of spontaneous emission and dephasing errors in both the ancilla and the cavity. Once prepared, these arbitrary logical states can be utilized via state injection to enable higher-level error correction or fault-tolerant gate implementations~\cite{zhou2000methodology,gottesman2009introduction,hillmann2022performance}, a direction we reserve for future work.

\begin{figure}[!t]
\begin{centering}
\includegraphics[scale=0.6]{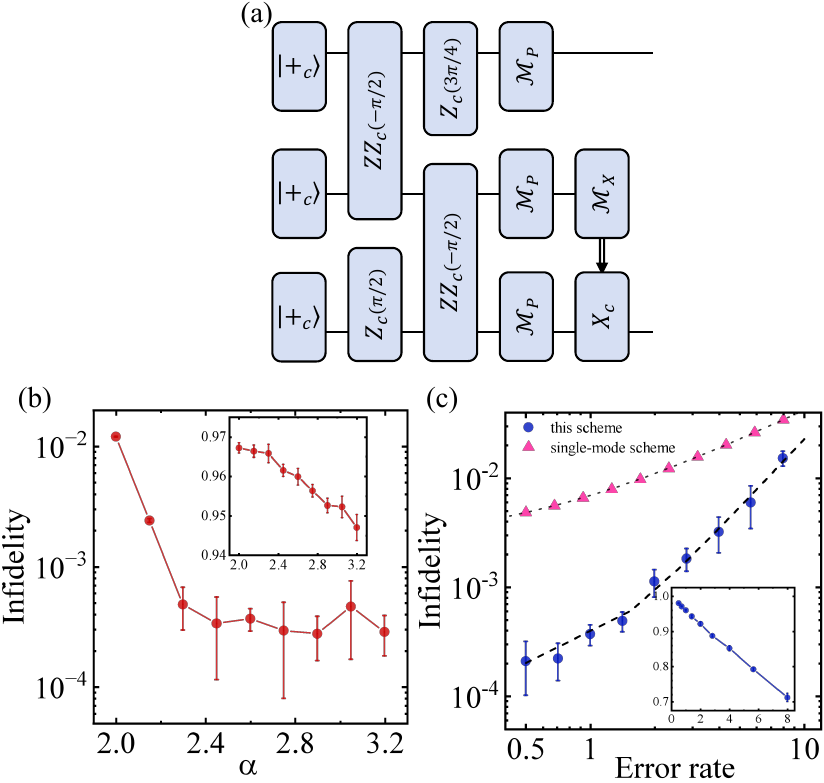}
\par\end{centering}
\caption{\label{Fig:BellStatePrep}Fault-tolerant preparation of the logical state $(\ket{0_\mathrm{c}}\ket{0_\mathrm{c}}+{e}^(i\pi/4)\ket{1_\mathrm{c}}\ket{1_\mathrm{c}})/\sqrt{2}$. ({a}) The corresponding quantum circuit for the preparation of the logical state, composed of the elementary gates shown in Fig.~\ref{Fig:UniversalCircuit}. Here, $\mathcal{M}_\mathrm{P}$ is the parity measurement post-selecting the even parity states, and $\mathcal{M}_\mathrm{X}$ is the Pauli-X measurement of the cat state. ({b}) Logical infidelity (main panel) and success probability (inset) for the preparation of the logical state as a function of the amplitude $\alpha$. ({c}) Logical infidelity (main panel) and success probability (inset) as a function of the physical error rate factor at $\alpha=2.6$. The blue dots represent the result from the new protocol, while the pink line is a baseline result of a single cavity and a single ancillary qubit initialized with SNAP gate. The black dashed line is a power-law fit to the blue points, given by $(1.6+1.8s^{2.12})\times10^{-4}$. The fitted scaling is greater than 1, which confirms the suppression of first-order noise. The parameters of the simulation are shown in the Supplemental Material. }
\label{Fig:BellStatePrep}
\end{figure}

\smallskip{}
\noindent\textit{Numerical results.-} We validate the protocol through numerical simulations on a realistic model of 3D circuit-QED platform in Fig.~\ref{Fig:Schematic}(c), using parameters from a recent experimental demonstration~\cite{ni2023beating}. Given the intractable dimension of the system's Hilbert space, all simulations employ an efficient Monte Carlo quantum trajectory algorithm~\cite{abdelhafez2019gradient,kuprov2007polynomially} that evolves the no-jump trajectory deterministically and only samples rare quantum jumps. All logical states are characterized via logical tomography and more details about the simulation are provided in Ref.~\cite{SM}.

As a first benchmark, we prepare the two-qubit entangled logical state $\left|\psi_\mathrm{L}\right\rangle = \ket{0_\mathrm{c}}\ket{0_\mathrm{c}} + e^{i\pi/4} \ket{1_\mathrm{c}}\ket{1_\mathrm{c}}$, which is a representative resource state relevant to magic-state injection. The corresponding quantum circuit, compiled from the basic gate set in Eq.~(\ref{Equ:UniversalGate}),  is shown in Fig.~\ref{Fig:BellStatePrep}(a). Three cavities are initialized in $\ket{+_{\mathrm{c}}}$. The subsequent $Z_c$ and $ZZ_c$ rotations, combined with $\mathcal{M}_X$-based teleportation, give rise to the FT preparation of this magic-state resource state.

Figure~\ref{Fig:BellStatePrep}(b) shows the logical infidelity and success probability as functions of the cat-code amplitude $\alpha\in [2.2,3.0]$. The infidelity initially decreases with $\alpha$ as the four coherent-state components become better separated, while it saturates at larger amplitudes as the photon decay error increases. This trade-off gives rise to a near-optimal logical infidelity $\overline{f}_L = (2.74\pm1.09)\times10^{-4}$ for $\alpha=2.9$, with a success probability exceeding $95\%$. Note that the conditional Pauli-$X_\mathrm{c}$ correction is implemented virtually using standard Pauli frame tracking~\cite{chamberland2018fault,knill2005quantum}. This avoids feedback latency and errors by absorbing corrections into the final measurement basis.

To verify the suppression of the first-order errors, we test the scaling of logical infidelity $\overline{f}_L$ against a global physical error. Here, we introduce an error multiplier factor $s$, i.e., scale all noise strengths in the whole system by a common factor $s$ ($\kappa \to s\kappa$) with a fixed $\alpha=2.6$. The results are shown in Fig.~\ref{Fig:BellStatePrep}(c). Our protocol exhibits a power-law scaling of $\overline{f}_L \propto s^{2.12\pm0.09}$. The exponent, significantly greater than 1, provides strong evidence that first-order errors are successfully suppressed. Meanwhile, the success probability exhibits a near-linear dependence on $s$, remaining above $70\%$ even at $s=8$.
For comparison, we simulate a conventional single-mode preparation, where the bosonic state $\left| 0_\mathrm{c}\right\rangle + e^{-i\pi/4} \left| 1_\mathrm{c}\right\rangle$ is prepared with the assistance of a two-level ancilla. The bosonic mode is initialized in a coherent state $\ket{\alpha}$, followed by a parity measurement and a selective number-dependent arbitrary phase (SNAP) gate~\cite{heeres2015cavity}. The simulation result shows the logical infidelity $\overline{f}_L \propto s^{0.961\pm0.003}$, confirming that such a traditional control approach cannot suppress first-order noise.

\begin{figure}[t]
\begin{centering}
\includegraphics[scale=0.6]{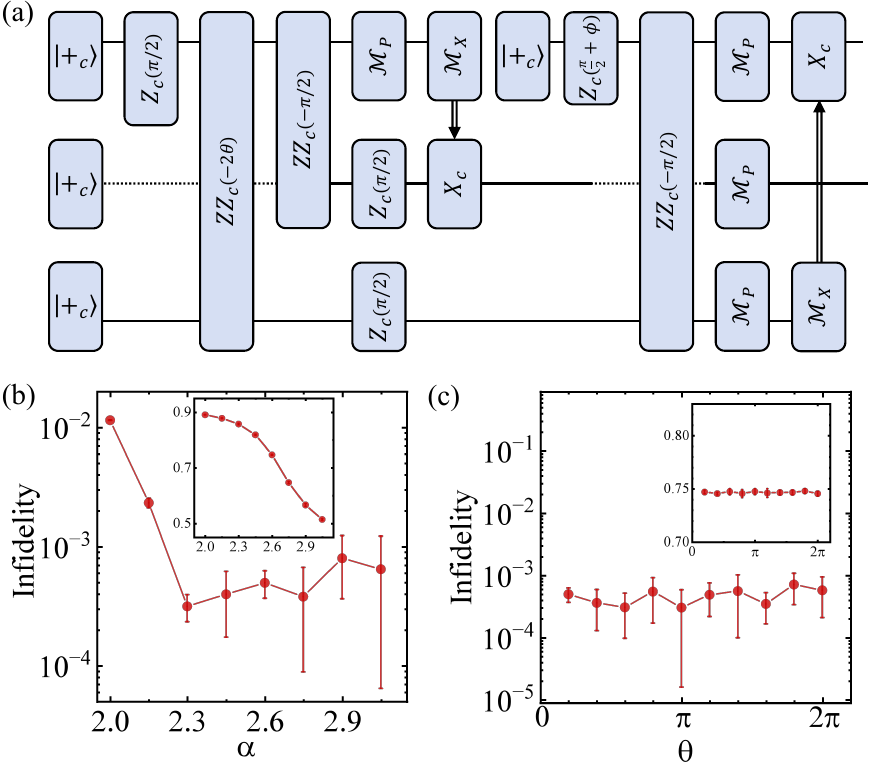}
\par\end{centering}
\caption{\label{Fig:ArbitaryState}{Fault-tolerant preparation of the logical state $\ket{\psi_\mathrm{L}(\theta,\phi)}=\mathrm{cos}\theta\ket{0_\mathrm{c}}\ket{0_\mathrm{c}}+ie^{i\phi}\mathrm{sin}\theta\ket{1_\mathrm{c}}\ket{1_\mathrm{c}}$}. ({a}) The quantum circuit constructed from the elementary operations shown in Eq.~(\ref{Equ:UniversalGate}). ({b}) Logical infidelity (main panel) and success probability (inset) for the logical state $\ket{\psi_\mathrm{L}(\pi/6,0)}$ as a function of the cat-code amplitude $\alpha$. ({c}) Logical infidelity (main panel) and success probability (inset) for the logical state $\ket{\psi_\mathrm{L}(\theta,0)}$ as a function of the parameter $\theta$, simulated at the amplitude $\alpha=2.6$. Other parameters of the simulation are shown in the Supplemental Material.}
\end{figure}

We further demonstrate the protocol's capability in arbitrary state preparation by preparing the continuous family of two-qubit logical states $\ket{\psi_\mathrm{L}(\theta,\phi)} = \mathrm{cos}\theta \ket{0_\mathrm{c}}\ket{0_\mathrm{c}} + ie^{i\phi}\mathrm{sin}\theta \ket{1_\mathrm{c}}\ket{1_\mathrm{c}}$ using a deeper circuit shown in Fig.~\ref{Fig:ArbitaryState}(a). The performance is also critically dependent on $\alpha$, as shown in Fig.~\ref{Fig:ArbitaryState}(b).
Specifically, the infidelity initially decreases and then increases with $\alpha$, reflecting the well-known trade-off between noise strength and the intrinsic distinguishability of the cat code~\cite{li2017cat}.
The near-optimal infidelity is found around $\alpha \in [2.2, 2.8]$. When $\alpha=2.6$, the logical infidelity is $\overline{f}_L=(5.0\pm1.3)\times10^{-4}$ , which is similar to the previous situation, showing that the logical infidelity is robust against the circuit depth.

To demonstrate versatility, we also simulate the preparation of different logical states and analyze their performance versus the phase $\theta$. The results in Fig.~\ref{Fig:ArbitaryState}(c) show that the logical infidelity is almost entirely independent of $\theta$. This stability arises because $\theta$ is set by the first $ZZ_\mathrm{c}(-2\theta)$-rotation gate whose duration is short relative to other fixed-time operations. This confirms our protocol can efficiently prepare a continuous family of logical states, which will be beneficial to the state-injection-based QEC scheme~\cite{hillmann2022performance,gottesman2009introduction}.

\noindent
\noindent\textit{Conclusion.-}
We have presented a universal set of error-detectable gates based on the four-legged cat code. This framework enables the preparation of arbitrary logical states while suppressing dominant errors, a key gadget for higher-level concatenated codes. Through detailed numerical verification, we have confirmed the protocol's ability to suppress first-order spontaneous emission and dephasing errors in both the bosonic mode and the ancilla. Using realistic parameters from current hardware platforms, our scheme achieves a logical state preparation infidelity on the order of $10^{-4}$ with a high success probability over $95\%$. This fault-tolerant protocol is compatible with the current experimental platform without Hamiltonian or dissipation engineering. Therefore, it can be immediately applied to resource state preparation (e.g., magic states) for concatenated codes or adapted for state injection-based error correction schemes~\cite{hillmann2022performance}. By efficiently converting dominant physical errors into detectable events~\cite{chang2025surface}, our work offers a path to significantly reduce the resource overhead required for scalable fault-tolerant quantum computation.

\bigskip{}
\noindent \textbf{Acknowledgements}
\begin{acknowledgments}
\noindent This work was funded by the National Natural Science Foundation of China (Grant Nos. 12547179, 92265210, 12550006, 92365301, 92565301, 92165209, 12574539), the Quantum Science and Technology-National Science and Technology Major Project (2021ZD0300200). This work is also supported by the Fundamental Research Funds for the Central Universities, the USTC Research Funds of the Double First-Class Initiative, the supercomputing system in the Supercomputing Center of USTC, and the USTC Center for Micro and Nanoscale Research and Fabrication.
\end{acknowledgments}

\end{document}